\documentclass[aps,prl,twocolumn,showpacs,superscriptaddress]{revtex4-2}  
\usepackage{graphicx}  
\usepackage{amssymb}   
\usepackage{amsmath}   
\usepackage[colorlinks,allcolors=blue]{hyperref}

\begin{document}
\title{Current cross-correlations in a quantum Hall collider at filling factor two}

\author{Edvin G. Idrisov}
\affiliation{Department of Physics and Materials Science, University of Luxembourg, Luxembourg}

\author{Ivan P. Levkivskyi}
\affiliation{Dropbox Ireland, One Park Place, Hatch Street Upper, Dublin, Ireland}

\author{Eugene V. Sukhorukov}
\affiliation{D\'epartement de Physique Th\'eorique, Universit\'e de Gen\`eve, CH-1211 Gen\`eve 4, Switzerland}

\author{Thomas L. Schmidt}

\affiliation{Department of Physics and Materials Science, University of Luxembourg, Luxembourg}
\date{\today}

\begin{abstract}
We use the non-equilibrium bosonization technique to study the effects of Coulomb interactions in mesoscopic electron colliders based on quantum Hall (QH) edge states at filing factor $\nu = 2$. The current cross-correlations and Fano factor, which carry the information about the exclusion statistics, are calculated. It is shown that both these quantities have a non-analytical dependence on the source transparency, which scales as $\log(1/T_s)$ at small $T_s \ll 1$. This is the consequence of electron-electron interactions in the outgoing non-equilibrium states of the collider.

\end{abstract}

\pacs{}
\maketitle

The progress in experimental techniques at the
nanoscale has allowed experimentalists to design
composite systems based on quantum Hall (QH) edge states~\cite{Ezawa,Glattli1,Dario2,Roussel}.
This development has already led to a better understanding of fundamental phenomena in mesoscopic physics, such as phase coherence~\cite{IvanDephasing, Ji,Neder,Tewari,Duprez2,Clerk,IdrisovDephasing}, charge and heat quantization~\cite{Pierre1,Pierre2,Furusaki,IdrisovSET,Pierre4,Mitali1,Mitali2}, as well as equilibration and relaxation~\cite{Pierre5,Pierre6,Pierre7,Chalker,IvanRelaxation,ArturRelaxation, Fujisawa,Meir,IdrisovAmmeter,Heiblum2020}. In particular, these novel architectures have started the field of electron quantum optics, where devices and phenomena known from optics have been recreated in the electronic domain.

One essential element of these experimental setups is a quantum point contact (QPC), i.e., a narrow channel between two electrically conducting systems~\cite{QPC}. In the context of electron quantum optics, such QPCs act as electron beam splitters and provide a platform for tunneling experiments in the presence of different kind of injection sources~\cite{Glattli2,Waintal}. Particular examples are electron analogs of the photonic Hong-Ou-Mandel, Hanbury Brown and Twiss, Mach-Zehnder and Fabry-Perot experiments~\cite{Ivanbook}. The current and shot noise through a QPC connecting integer and fractional QH edge states have been measured in several experiments~\cite{Glattli1}. These studies have made it possible to distinguish Fermi liquid and non-Fermi liquid phases in current-voltage characteristics and in the corresponding current noise~\cite{Wen2,Chamon1,Chamon2,Martin1,Martin2,Chang,Kane1,Kane2,Dario1,Dario3,Edvin2}. Apart from that, it is worth mentioning that the noise
provides one of the most straightforward methods to measure the effective charge of tunneling quasiparticles in the fractional QH regime~\cite{Glattli2019}.

Using a combination of QPCs, experimentalists were able recently to build \emph{anyonic} colliders at a filling factor $\nu=1/m$ (where $m$ is an odd integer)~\cite{Gwendal2020}. Their measurements of the current cross-correlations at zero frequency and the generalized Fano factor have provided evidence for anyonic exchange statistics of charge carriers with an exchange phase $\pi/3$ for $m=3$. This agrees with the theoretical predictions for the Laughlin state at $\nu=1/3$~\cite{Rosenow}. It is worth mentioning that the measurement of the noise in such colliders is an alternative method for the investigation of quasiparticle statistics, compared to the conventional Mach-Zehnder and Fabry-Perot interferometers, where the measured current or conductance is sensitive to the exchange phase~\cite{Nakamura2020}.

\begin{figure}
\includegraphics[width=\columnwidth]{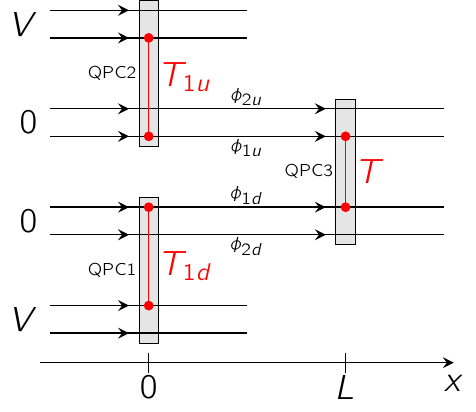}
\caption{\label{fig:one} Schematic picture of a mesoscopic electron collider based on QH edge states at filling factor $\nu=2$. Two independent voltage-biased sources QPC1 and QPC2 with respective transparencies $T_{1d}$ and $T_{1u}$ generate non-equilibrium states at the upper and lower edges, which collide at QPC3 with transparency $T$. The zero-frequency current noise $j_{1u}(t)$ and $j_{1d}(t)$ after QPC3 is measured. The charge density fields are expressed in terms of non-equilibrium chiral boson fields $\phi_{\alpha s}(x,t)$ where $\alpha \in \{1,2\}$ and $s\in \{u,d\}$.}
\end{figure}

In this paper, we consider a mesoscopic electron collider based on QH edge states at integer filling factor $\nu=2$, where the incoming non-equilibrium states are created using two quantum point contacts (QPC1 and QPC2), with respective transparencies $T_{1s} \ll 1$, where $s \in \{u,d\}$ denotes the upper and lower channels (see Fig.~\ref{fig:one} for a schematic illustration). At zero temperature, the resulting incoming states have ``double-step'' distribution functions $f_{1s}(\epsilon)=\theta(-\epsilon)+T_{1s} \theta(\epsilon) \theta(V-\epsilon)$, where $V$ is the source voltage. Due to Coulomb interactions the incoming states split into two modes with different velocities. Subsequently, the non-equilibrium modes arriving from the upper and lower part of the collider are mixed by a third QPC (QPC3) with transparency $T$. In order to investigate the statistical fluctuations after the collision at QPC3, which yield information about the exclusion statistics, we calculate the zero-frequency cross-correlations of the current~\cite{Gwendal2020}
\begin{equation}
\label{Definition of cross noise}	
S_{1u;1d}(\omega=0)=2 \int dt \langle \delta j_{1u}(t) \delta j_{1d}(0) \rangle,
\end{equation}
where $\delta j_{1s}(t)=j_{1s}(t)-\langle j_{1s}(t)\rangle$ and the average is taken with respect to non-equilibrium state created by injection currents from QPC1 and QPC2. This quantity has been measured in recent experiment~\cite{Gwendal2020}. For chiral fermions the current in Eq.~(\ref{Definition of cross noise}) is proportional to the charge density, $j_{1s}(x,t) \propto \psi^{\dagger}_{1s}(x,t) \psi_{1s}(x,t)$. Apart from the current noise, we are interested in the generalized Fano factor, which contains information about the exclusion statistics~\cite{Gwendal2020}. It is defined as
\begin{equation}
\label{Definition of Fano factor}
P=\frac{S_{1u;1d}(0)}{2eT(1-T) I_{+}},
\end{equation}
where $I_{+}=e^2 V (T_{1u}+T_{1d})/2\pi \hbar$ is the total input current. The theoretical study of current cross-correlations is non-trivial since one has to deal with a non-equilibrium situation due to the injection QPCs as well as with strong electron-electron interactions. For this reason, it is convenient to use the non-equilibrium bosonization approach~\cite{IvanNEB1,IvanNEB2,Gutman} which we explain below. Throughout the paper, we focus on the case of zero temperature. We set $|e|=\hbar=k_B=1$ during the calculation and restore the dimensions in main results for the noise at the end.

\textit{Theoretical method}. To study quantum Hall edge states at filling factor $\nu=2$, it is convenient to use their effective field theory~\cite{Wen1}. According to this, the low-energy degrees of freedom of the edge states are collective
fluctuations of the charge densities $\rho_{\alpha s}(x)$, where the subscripts $\alpha \in \{1,2\}$ and $s \in \{u,d\}$ denote the channels (see Fig.~\ref{fig:one}). The  fermionic fields $\psi_{\alpha s}(x,t) \propto \exp[i\phi_{\alpha s}(x,t)]$ and the charge density fields, $\rho_{\alpha s}(x,t)=(1/2\pi) \partial_x \phi_{\alpha s}(x,t)$, are represented in terms of chiral boson fields which satisfy the standard equal-time commutation relations
\begin{equation}
\label{Commutation relation for bosonic fields}
[\phi_{\alpha s}(x,t),\phi_{\beta r}(y,t)]=i\pi \delta_{\alpha \beta} \delta_{sr} \text{sgn}(x-y).
\end{equation}
The Hamiltonian of the edge states in the presence of Coulomb interactions is given by
\begin{equation}
\label{Hamiltonian of edge states}
H=\frac{1}{2} \sum_{s, \alpha \beta}\int \int dx  dy V_{\alpha \beta}(x-y) \rho_{\alpha s}(x) \rho_{\beta s}(y),
\end{equation}
where the integral kernel $V_{\alpha \beta}(x-y)=2\pi v_F \delta_{\alpha \beta} \delta(x-y)+U_{\alpha \beta}(x-y)$ includes both the kinetic energy of the free chiral electrons with Fermi velocity $v_F$ as well as the Coulomb interaction between electrons with interaction potential $U_{\alpha\beta}(x)$. At low energies the characteristic length scales are much longer than the screening length of the Coulomb interaction. Therefore, one can further approximate the Coulomb potential as short-ranged, i.e., $U_{\alpha \beta}(x-y)=U_{\alpha \beta}\delta(x-y)$, with non-zero diagonal and off-diagonal interaction strengths $U_{\alpha\beta}$~\cite{Ivanbook}.

The current cross-correlations can be calculated by using the scattering approach locally at QPC3~\cite{Blanter}. One finds that the cross-correlations of the outgoing currents after QPC3 at zero frequency are given by a sum of two terms~\cite{Rosenow,Gwendal2020}
\begin{equation}
\label{Total cross noise}
S_{1u;1d}(\omega=0)=\mathcal{P}_1(0)+\mathcal{P}_2(0).
\end{equation}
The first term is associated with a transmitted noise and has the simple form,
\begin{equation}
\label{Transmitted noise}
\mathcal{P}_1(0)=\frac{e^2 V}{2\pi \hbar} (1-T)T\sum_{s=u,d}T_{1s}(1-T_{1s}).
\end{equation}
The second term describes the generated noise by QPC3,
\begin{equation}
\label{Generated noises}
\mathcal{P}_2(0)=2(1-T)T a^{-2}\int dt K_u(t) G_d(t).
\end{equation}
where $a$ is the short-distance cut-off from bosonization and
\begin{equation}
\label{Correlation function}
\begin{aligned}
K_u(t)= \langle e^{-i\phi_{1u}(L,t)} e^{i\phi_{1u}(L,0)}\rangle, \\
G_d(t)= \langle e^{i\phi_{1d}(L,t)} e^{-i\phi_{1d}(L,0)} \rangle.
\end{aligned}
\end{equation}
All bosonic fields should be evaluated at the position $x = L$ of QPC3 and the averages in the correlation functions are taken with respect to the non-equilibrium state created by the source QPC1 and QPC2. Next, we apply the non-equilibrium bosonization technique~\cite{IvanNEB1,IvanNEB2,Gutman} to evaluate these correlation functions.

The Hamiltonian of the edge states in Eq.~(\ref{Hamiltonian of edge states}) generates the equation of motion of the bosonic fields $\phi_{\alpha s}(x,t)$. The latter have to be supplemented with the appropriate boundary conditions due to source QPCs, resulting in
\begin{align}
\label{Equation of motions with boundary conditions}
\partial_t \phi_{\alpha s}(x,t)&=-\frac{1}{2\pi} \sum_{\beta} \int dy V_{\alpha \beta}(x-y) \partial_y \phi_{\beta s}(y,t), \notag \\
\partial_t \phi_{\alpha s}(0,t)&=-2\pi j^{{\rm in}}_{\alpha s}(t),
\end{align}
where the incoming currents $j^{{\rm in}}_{\alpha s}(t)$ are defined at the point $x=0^+$ right after the source QPC1 and QPC2. These equations of motion can
be solved by diagonalizing the interaction matrix $V_{\alpha \beta }=2\pi v_F \delta_{\alpha \beta}+U_{\alpha \beta}$ via a Bogoliubov
rotation, $V=S(\theta) \Lambda S^{\dagger}(\theta)$, which conserves the commutation relations of the bosonic fields. The rotation angle is determined by the strength of the electron-electron interaction and is given by $\tan (2\theta)=2V_{12}/(V_{11}-V_{22})$. The new collective excitations are characterized by two velocities, $\Lambda=\text{diag}(u,v)$, where $u,v=(V_{11}+V_{22})/2 \pm \sqrt{(V_{11}-V_{22})^2/4+V^2_{12}}$ and we used that $V_{12}=V_{21}$. Imposing the boundary conditions~(\ref{Equation of motions with boundary conditions}) we obtain the dynamics of bosonic fields in both channels $s \in \{u,d\}$,
\begin{align}
\label{Solution of equation of motion}
    \phi_{1s}(x,t)
&=
    -\lambda_1 Q_{1s}(t_u)-\lambda_2 Q_{2s}(t_u) \notag \\
&
    -\lambda^{\prime}_1 Q_{1s}(t_v)+\lambda_2 Q_{2s}(t_v),  \\
    \lambda_1
&=
    \pi[1-\cos(2\theta)], \notag \\
    \lambda_2
&=
    \pi \sin(2\theta), \notag \\
    \lambda^{\prime}_1
&=
    2\pi-\lambda_1,\notag
\end{align}
where we have introduced the injected charges
\begin{align}
    Q_{\alpha s}(t)=\int^t_{-\infty} dt^{\prime} j^{{\rm in}}_{\alpha s}(t^{\prime})
\end{align}
and the times $t_u=t-x/u$ and $t_v=t-x/v$. Therefore, the bosonic fields~(\ref{Solution of equation of motion}) at observation point $x$ at time $t$ are determined by the charges which arrive with different delay times $t_u$ and $t_v$, and these charges are weighted with coupling constants due to intra- and inter-channel electron-electron interactions. Now we have all necessary ingredients to calculate the generated noise from Eq.~(\ref{Generated noises}).

\textit{Gaussian noise regime}.
First, we consider the case of Gaussian fluctuations of the bosonic fields. This means that the correlation function in Eq.~(\ref{Correlation function}) can be written as $\log K_u(t)=2\pi \langle j^{{\rm in}}_{1u}\rangle t+\langle \phi_{1u}(L,t)\phi_{1u}(L,0)\rangle-\langle \phi^2_{1u}(L,t)\rangle/2-\langle \phi^2_{1u}(L,0)\rangle/2$, where $\langle j^{{\rm in}}_{1u} \rangle=VT_{1u}/2\pi$ is the current injected from upper QPC1. Now using Eq.~(\ref{Solution of equation of motion}) we obtain
\begin{equation}
\label{Correlation function equilibrium}
\begin{aligned}
& \log K_{u}(t)=iT_{1u} V t-\int \frac{d\omega}{\omega^2} (1-e^{-i\omega t})\left( \mathcal{F}_{1}+\mathcal{F}_{2}\right),\\
& \mathcal{F}_{1}=2\pi\left[1-\frac{\lambda_1 \lambda^{\prime}_1}{\pi^2}\sin^2\left( \frac{\omega t_L}{2}\right)\right]S^{{\rm in}}_{1u}(\omega), \\
&\mathcal{F}_{2}=\frac{2\lambda^2_2}{\pi}\sin^2\left( \frac{\omega t_L}{2}\right) S^{{\rm in}}_{2u}(\omega),
\end{aligned}
\end{equation}
where we have introduced the spectral function of current fluctuations, $S^{{\rm in}}_{1u}(\omega)=\int dt e^{i\omega t} \langle \delta j^{{\rm in}}_{1u}(t) \delta j^{{\rm in}}_{1u}(0)\rangle$, and the delay time between wave packets, $t_L=L/v-L/u$, where $L$ is the distance between the injection QPCs and QPC3. A similar expression can be obtained for $G_d(t)$ by replacing the subscript $u$ with $d$.

To obtain the spectral function of the incoming state, we again use the free-fermion scattering approach~\cite{Blanter} because the interactions do not influence the tunneling at the injection QPCs. The general result for $\alpha \in \{1,2\}$ and $s \in \{u,d\}$ is then given by the sum of two terms,
\begin{equation}
\label{Equilibrium noise}	
S^{{\rm in}}_{\alpha s}(\omega)=S_q(\omega)+(1-T_{\alpha s}) T_{\alpha s} S_{n}(\omega),
\end{equation}
where $S_q(\omega)=\omega \theta(\omega)/2\pi$ is the equilibrium ground state contribution and $S_n(\omega)=\sum_{\pm} S_q(\omega \pm V)-2S_q(\omega)$ is the non-equilibrium part which originates from the voltage-biased injection QPCs. Substituting Eq.~(\ref{Correlation function equilibrium}) into Eq.~(\ref{Generated noises}), the second term in Eq.~(\ref{Total cross noise}) takes the following final form
\begin{equation}
\mathcal{P}_2(0)=-\frac{e^2 V}{2\pi \hbar} 2(1-T)T\int \frac{dz}{2\pi} \frac{e^{i(T_{1u}-T_{1d}) z}}{(z-i0^+)^2} K_n(z),
\end{equation}
where we have introduced the dimensionless integration variable $z=Vt$ and used
\begin{align}
& \log K_n(z)=-2\left[(1-T_{1u}) T_{1u}+(1-T_{1d}) T_{1d}\right] \\
&\times \int_0^1 \frac{dx}{x^2} (1-x)[1-\cos(z x)]\left[1-\frac{\lambda_1 \lambda^{\prime}_1}{\pi^2}\sin^2\left(\frac{ x L}{2 L_{ex}} \right)\right], \notag
\end{align}
where $L_{ex}=uv/[(u-v)V]$ is the characteristic correlation length between the two modes with velocities $u$ and $v$ and $x=\omega/V$ is the dimensionless variable of integration.
\begin{figure}
\includegraphics[width=\columnwidth]{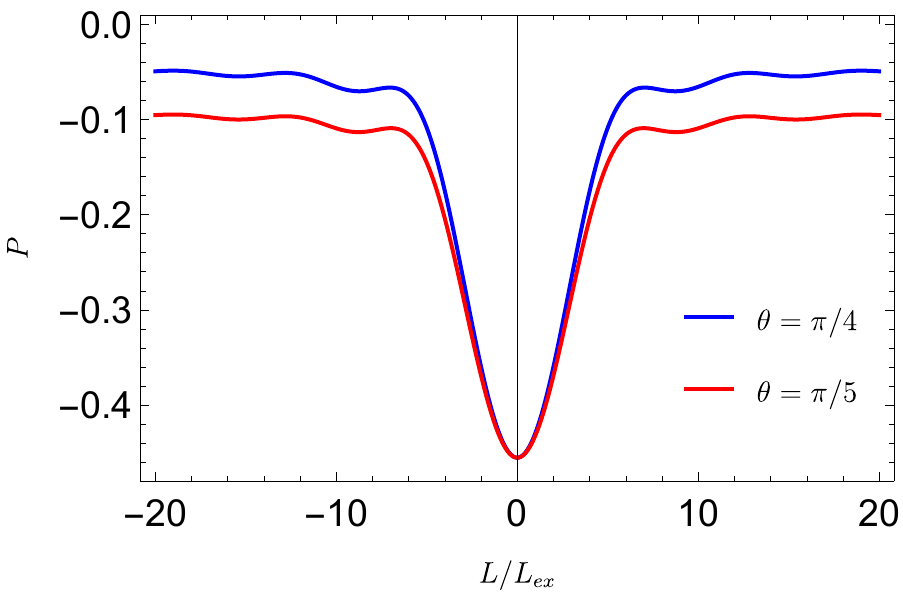}
\caption{\label{fig:two} The generalized Fano factor $P$ as a function of the dimensionless length $L/L_{ex}$ for different values of the interaction strength, parameterized by $\lambda_1=\pi[1-\cos(2\theta)]$, in the Gaussian noise regime. We set $T_s=0.15$.}
\end{figure}

In order to understand and compare the case of Gaussian noise to experimental data, we consider the generalized Fano factor~(\ref{Definition of Fano factor}) at $T_{1u}=T_{1d}=:T_s$,
\begin{equation}
\begin{aligned}	
& P=1-T_s+\frac{1}{T_s} \int \frac{dz}{2\pi} \frac{e^{-4 R_s T_s  \Phi(\lambda_1, L/L_{ex}, z)}}{(z-i0^+)^2}, \\
& \Phi(\lambda_1, L/L_{ex}, z)=\\
& \int_0^1 \frac{dx}{x^2} (1-x)[1-\cos(z x)]\left[1-\frac{\lambda_1 \lambda^{\prime}_1}{\pi^2}\sin^2\left(\frac{ x L}{2 L_{ex}} \right)\right],
\end{aligned}
\end{equation}
as a function of the dimensionless variable $L/L_{ex}$. Figure~\ref{fig:two} shows that the Fano factor is negative, in contrast to experimental data from Ref.~\cite{Gwendal2020}, indicating that the experiment was perhaps not performed in the Gaussian regime.

Finally, we would like to mention briefly that our results do not apply in the strictly non-interacting limit. In that case, the two modes have identical velocities, $u=v$, which leads to a divergence $L_{ex} \to \infty$.

\textit{Non-Gaussian noise regime}.
Next, we consider the fluctuations of the bosonic fields as non-Gaussian at long distances $L \gg L_{ex}/\text{min}\{T_{1u},T_{1d}\}$. At such distances, the partitioned charges $Q_{1s}$ at the source QPCs, taken at times $t_u$ and $t_v$, can be considered as uncorrelated. Therefore,  the correlation functions associated with the upper and lower parts of the collider in Eq.~(\ref{Correlation function}) split into products of four factors
\begin{equation}
\label{Ku and Gd in terms of FCS four terms}
\begin{aligned}
K_u(t)=\chi_{1u}(\lambda_1, t) \chi_{1u}(\lambda^{\prime}_1,t)\chi_{2u}(\lambda_2,t)\chi_{2u}(-\lambda_2,t),\\
G_d(t)=\chi_{1d}(-\lambda_1, t) \chi_{1d}(-\lambda^{\prime}_1)\chi_{2d}(-\lambda_2,t)\chi_{2d}(\lambda_2,t),
\end{aligned}
\end{equation}
where each factor represents a generator of full counting statistics (FCS)~\cite{Ivanbook},
\begin{equation}
\chi_{\alpha s}(\lambda, t)=\langle e^{i\lambda Q_{\alpha s}(t)} e^{-i\lambda Q_{\alpha s}(0)}\rangle.
\end{equation}

For the non-equilibrium state created at QPC1 and QPC2, the generating function can be written analytically in the long-time Markovian limit, which gives the main contribution to integrals in Eq.~(\ref{Generated noises}). This is known as the Levitov-Lee-Lesovik formula~\cite{Levitov}. In the Poissonian regime of small transparencies $T_{1s} \ll 1$, the Levitov-Lee-Lesovik formula for positive times $t>0$ is given by
\begin{equation}
\label{Levitov Lesovik formula}
\log \chi_{\alpha s}(\lambda, t)=\frac{\lambda^2}{4\pi^2}\log\left(\frac{\gamma}{it+\gamma}\right)-\frac{Vt}{2\pi} T_{\alpha s}(1-e^{i\lambda}).
\end{equation}
where $\gamma \propto a/(2 \pi v_F)$ is related to the short-distance cutoff $a$. For $t < 0$ the FCS can be calculated by analytic continuation of Eq.~(\ref{Levitov Lesovik formula}), namely $\chi_{\alpha s}(\lambda, -t)=\chi^{\ast}_{\alpha s}(\lambda, t)$. Since the channels with subscript $\alpha=2$ are not biased and $T_{2s}=0$, the FCS generators for these channels are given by the first term of Eq.~(\ref{Levitov Lesovik formula}). It is worth mentioning that this is the ground state (Fermi sea) contribution. However, the channels with $\alpha=1$ include the second term, which is a non-Gaussian contribution.
\begin{figure}
\includegraphics[width=\columnwidth]{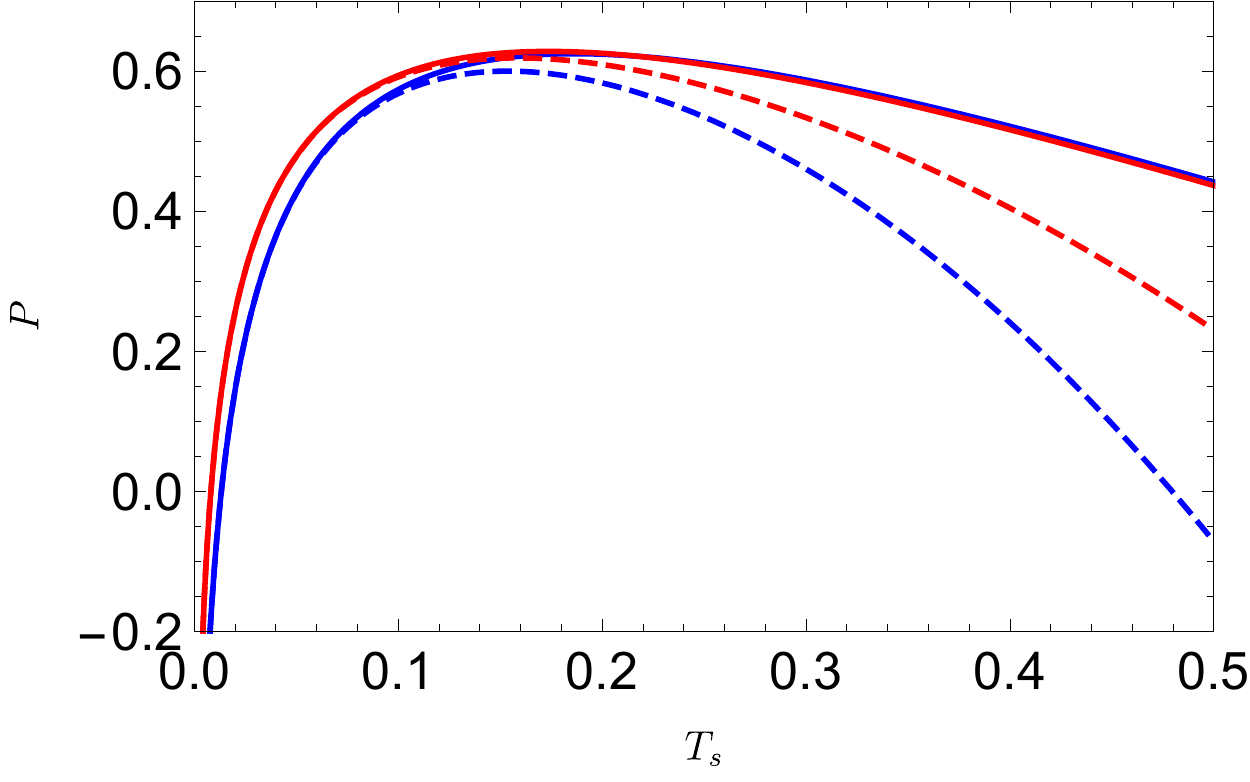}
\caption{\label{fig:three} The generalized Fano factor $P$ as a function of source transparency $T_s$ in the regime of non-Gaussian noise. The solid lines are numerical results based on Eq.~(\ref{Definition of Fano factor}) and the dashed lines are the asymptotics in Eq.~(\ref{Fano factor. NonGaussian regime}). The blue and red curves correspond to $\theta=\pi/4$ (strong interaction) and $\theta=\pi/5$.}
\end{figure}
Substituting the FCS generator~(\ref{Levitov Lesovik formula}) into Eqs.~(\ref{Ku and Gd in terms of FCS four terms}) and (\ref{Generated noises}), the second term of the noise at $T_{1u}=T_{1d}=:T_s$ turns out to be equal to
\begin{equation}
\label{Generated noise in non-Gaussian case}
\mathcal{P}_2(0)=-\frac{e^2 V}{2\pi \hbar} \int \frac{dz}{2\pi} \frac{\exp\left[-2 T_s f(\lambda_1) |z|/\pi \right]}{(z-i0^+)^2},
\end{equation}
where $f(\lambda_1)=1-\cos(\lambda_1)$ and $z=Vt$ is the dimensionless integration variable. Now, substituting Eq.~(\ref{Generated noise in non-Gaussian case}) into Eq.~(\ref{Total cross noise}) and then into Eq.~(\ref{Definition of Fano factor}), we get the final formula for the generalized Fano factor to leading order in the transparency of the injection QPCs
\begin{equation}
\label{Fano factor. NonGaussian regime}	
\begin{aligned}	
& P=1-T_s+\frac{2 f(\lambda_1)}{\pi^2} \left[ \mathcal{C}-\pi \tilde{T}_s -\log(1/\tilde{T}_s) \right],
\end{aligned}
\end{equation}
where $\mathcal{C}=\log 2+\boldsymbol{\gamma}$, $\boldsymbol{\gamma} \approx 0.5772$ is Euler's constant, and $\tilde{T}_s=T_s f(\lambda_1)$ is the renormalized transparency of the source QPCs. Note that for $\theta=\pi/4$, which corresponds to strong electron-electron interactions, one finds $f(\lambda_1)=2$. It is worth mentioning that the subleading corrections in $T_s$ show non-analytical behavior as well and are proportional to $\log(1/\tilde{T}^2_s)$. The result~(\ref{Fano factor. NonGaussian regime}) does not depend on length explicitly because the calculations were done in the limit of large $L$, when a complete separation of the wave packets with velocities $u$ and $v$ occurs. The generalized Fano factor as a function of source transparency $T_s$ and for different interaction strengths $\theta$ is plotted in Fig.~\ref{fig:three}. 

To summarize, we have studied the zero-frequency current cross-correlations and the corresponding generalized Fano factor in a mesoscopic quantum Hall collider at filling factor $\nu=2$. We have used non-equilibrium bosonization in order to take into account both the non-equilibrium state created by the source QPCs as well as the electron-electron interactions. We have shown that both the noise and the Fano factor consist of two terms. The first term is associated with the transmitted noise and is proportional to $T_s(1-T_s)$. This term only contains information about the non-equilibrium conditions after the injection QPCs. The second term, which mixes the incoming non-equilibrium states, is related with the generated noise. It shows a non-analytical dependence on the source transparency and scales as $T_s \log(1/T_s)$ at small $T_s \ll 1$ in case of strong, screened Coulomb interactions ($\theta=\pi/4$). The non-analytical behavior is robust with respect to the interaction parameter and does not vanish for $\theta \neq \pi/4$. Thus, the generated noise contains information both about the non-equilibrium state after the source QPCs and the electron-electron interactions. The same considerations are valid for the generalized Fano factor. It is known that the absence of current correlations
for free fermions at filling factor $\nu=1$ is due to perfect spatial exclusion, while the negative correlations for Laughlin anyons at $\nu=1$ in non-Gaussian noise regime are a signature of a reduced spatial exclusion~\cite{Rosenow}.
However, for filling factor $\nu=2$ and experimentally accessible transparencies $T_s \geq 0.05$~\cite{Gwendal2020}, we have shown that the Fano factor in the non-Gaussian regime of the injection QPCs is positive, in contrast to the cases for free fermions and Laughlin quasiparticles. The positive correlations arise due to the additional Coulomb interactions between fermions.

We are grateful to G. F\`eve for sharing experimental data and for fruitful discussions. EGI and TLS acknowledge financial support from the National Research Fund Luxembourg under Grant CORE 13579612. EVS acknowledges support from the Swiss National Science Foundation.

\bibliography{refs}

\end{document}